\documentclass[aps,pre,reprint,superscriptaddress]{revtex4-2}

\usepackage{graphicx}
\usepackage{xcolor}
\usepackage{lipsum}   

\begin{document}
\title{Quadrilateral Particle Arrangement within Shocks in a Two-Dimensional Dusty Plasma}
\author{Anton\ Kananovich}
\affiliation{Department of Physics and Astronomy, Appalachian State University, Boone, North Carolina 28608, USA}
\affiliation{Department of Physics and Astronomy, University of Iowa, Iowa City, Iowa 52242, USA}
\author{J.\ Goree}
\affiliation{Department of Physics and Astronomy, University of Iowa, Iowa City, Iowa 52242, USA}
\date{\today}
\begin{abstract}
The microscopic structure within a two-dimensional shock was studied using data from a dusty plasma experiment. A single layer of charged microparticles, levitated in a glow-discharge plasma, was perturbed by an electrically floating wire that was moved at a steady supersonic speed to excite a compressional shock. A rearrangement of particles was observed, from a hexagonal lattice in the preshock into a quadrilateral microstructure within the shock. This quadrilateral structure would not be stable in a monolayer of identical repulsive particles, under equilibrium conditions. Glaser-Clark polygon analysis of the microstructure helped in identifying quadrilaterals.  Voronoi analysis was used to characterize the defect fraction behind the shock, as an indication of shock-induced melting.
\end{abstract}
\maketitle

\section{\label{secIntro}Introduction}

Shock waves (shocks), which can be defined as disturbances propagating at supersonic speed, exist in all kinds of media, including gases, solids, and plasmas. One type of plasma, which can also sustain shocks, is two-dimensional (2D) dusty plasma. This type of plasma consists of a layer of microparticles levitating in the sheath above the lower electrode in the gas discharge~\cite{goree2008fundamentals,melzer2019physics}. The ground state for such a 2D layer is a triangular lattice with six-fold symmetry.

Shocks in 2D dusty plasma have recently seen an increase in interest, including experimental and molecular dynamics (MD) simulations studies~\cite{samsonov1999mach,samsonov2000mach,samsonov2004shock,couedel2012three,lin2019pressure,kananovich2020shocksa,kananovich2020experimental,lin2020universal,tao2020effects,sun2021shock,qiu2021observation,kananovich2021shock,sun2021evolution,ding2021head, gou2021effects,li2021weak,tao2021effect,zhang2022structural,samsonov1999mach,samsonov2000mach,samsonov2004shock,couedel2012three,lin2019pressure,lin2020universal,tao2020effects,sun2021shock,qiu2021observation,sun2021evolution,ding2021head,gou2021effects,li2021weak,tao2021effect,zhang2022structural}. Among recently reported MD simulations, we can mention studies of shock-front speed and hydrodynamic quantities~\cite{lin2019pressure}, hydrodynamic quantities and velocity distribution functions along with analytical expression between the thermal and the drift velocities~\cite{lin2020universal}, the evolution of blast waves~\cite{sun2021evolution}, and a head-on collision of shocks~\cite{ding2021head}.

Due to its unique properties, 2D dusty plasma is an excellent medium for experimentally studying the microscopic structure of shocks. The sound speed in that medium is the order of centimeters per second so that video microscopy diagnostics allow the resolving and tracking of individual microparticles~\cite{samsonov2008high}. These particle-level measurements allow experimental study of phenomena such as the finite thickness of the shock front~\cite{kananovich2021shock}, and they also allow direct comparisons to MD simulations. The results may also be applicable to other 2D physical systems that similarly exhibit an equilibrium state with hexagonal structure~\cite{zhang2022structural}.

The microscopic properties of shocks in 2D dusty plasma have been studied using the method of Voronoi diagrams, with an input of particle-position data obtained from MD simulations. Sun and Feng~\cite{sun2021shock} studied melting, which was reported to occur when the moving exciter had a speed exceeding a threshold, as determined by Voronoi diagrams that were examined qualitatively and were also used to quantify defect fraction. A leap of the defect ratio, which can indicate the occurrence of a phase transition from the solid to liquid states, was observed. Qiu~et~al.~\cite{qiu2021observation} studied a postshock region of 2D crystal, with both liquid and solid coexisting phases, which was observed when the exciter speed was sufficiently high. For this phenomenon, again Voronoi diagrams were used to examine disorder and quantify defect fraction.

In this paper, we report an experimental investigation of a point that has not been much studied yet: the microscopic structure at the shock front and immediately behind it. We find that the six-fold triangular geometry of a ground-state crystal does not persist, under the sudden compression of a shock, but instead the particles tend to rearrange into a quadrilateral microscopic structure. The quadrilateral microstructure is unusual, because for a 2D crystal with isotropic repulsion the triangular (also called hexagonal) lattice is the only close-packed lattice in two dimensions~\cite{chaikin1995principles}, under the conditions of steady equilibrium. A quadrilateral lattice in 2D dusty  plasma under steady conditions can exist only in special cases, such as binary mixtures~\cite{huang2019wave} or when the crystal buckles~\cite{zampetaki2020buckling,singh2022square}. In this paper, we observe quadrilateral microscopic structure under the conditions of shock compression that are not a steady equilibrium.

To study this microscopic structure, we analyze our experimental particle-position data using the methods of Voronoi diagrams~\cite{voronoi1908nouvelles} and the polygon construction method of Glaser and Clark~\cite{ruhunusiri2011polygon,glaser1990statistical,glaser1992melting}. The polygon construction method is particularly effective in revealing the quadrilateral microscopic structure that we report, for the nonequilibrium conditions of shock compression in a 2D substance.

\section{Method}
\label{sec_method}
\subsection{Experiment}
In this paper, we report a further analysis of experimental runs that we previously reported Ref.~\cite{kananovich2020experimental}. In that experiment, as an exciter was moved at a steady speed, there was a localized compression of the 2D dust layer in front of the exciter. This compression was strong and supersonic, indicating that it was a shock. The  main result in that previous report was the relationship of two speeds, for the shock itself and for the exciter. Here we analyze the particle-position data for a different purpose, to assess the microscopic structure, particularly within the shock and immediately behind it. We briefly review that experiment next; further details are provided in Ref.~\cite{kananovich2020experimental}.

The experiment was performed in our Kuda-Topf chamber~\cite{kananovich2020experimental,ruhunusiri2014investigation,chaubey2021positive}. To sustain a gas discharge in argon, 13.56 MHz radio-frequency (rf) power with a peak-to-peak voltage of 184 V was applied to the lower electrode through a coupling capacitor, while the outer wall of the chamber was grounded. As a result, a dc bias of -85.5 V developed on the lower electrode. A 2D layer of monodisperse (8.69 \textmu m in diameter) melamine-formaldehyde microparticles was levitated in the plasma sheath above the lower electrode. The particle charge was $-1.4 \times 10^4$~elementary charges. In the absence of the perturbation of a shock, the lattice constant was $b = 0.80$~mm, and correspondingly the 2D Wigner-Seitz radius was $a = 0.42$~mm, while the nominal 2D dust plasma frequency was $\omega_{pd} = 49$~s\textsuperscript{-1}.

Compressional shocks were generated by continuously moving an exciter at a controlled velocity. The exciter was a nickel wire, shown in the Supplemental Material~\footnote{See Supplemental Material at [URL will be inserted by publisher]}. This exciter wire was aligned perpendicular to its direction of motion but parallel to the 2D microparticle cloud. Runs were performed at five different speeds of the exciter at various speeds, ranging from 50.8 to 101.6~mm/s  all above the sound speed of~19 mm/s. In dimensionless units, this range was from 2.5 to 5.0~$a \omega_{pd}$, significantly exceeding the exciter speeds in the simulation of~\citeauthor{sun2021evolution}~\cite{sun2021evolution}. 

For this paper, we analyzed the microscopic structure for all of these runs; we report snapshots of the data for three runs in the main text, and more detailed video data for all the runs in the Supplemental Material~\cite{Note1}.

In the experiment, the 2D microparticle cloud was imaged at 870~frames/s using video microscopy~\cite{feng2007accurate,feng2011errors,feng2016particle}. Afterwards, positions of individual microparticles were measured in each video frame, using the moment method, which has a sub-pixel precision~\cite{feng2007accurate,schneider2012nih}. We restricted our analysis to particles within a $25.4 \times 10.2$~mm region of interest (ROI), as described in Supplemental Material of Ref.~\cite{kananovich2020experimental}. The microparticle positions within the ROI, in each video frame, served as the input for the analysis that we describe next.

\subsection{Polylgon  analysis}

\begin{figure*}[ht]
	\centering
	\includegraphics[width=1.0\textwidth]{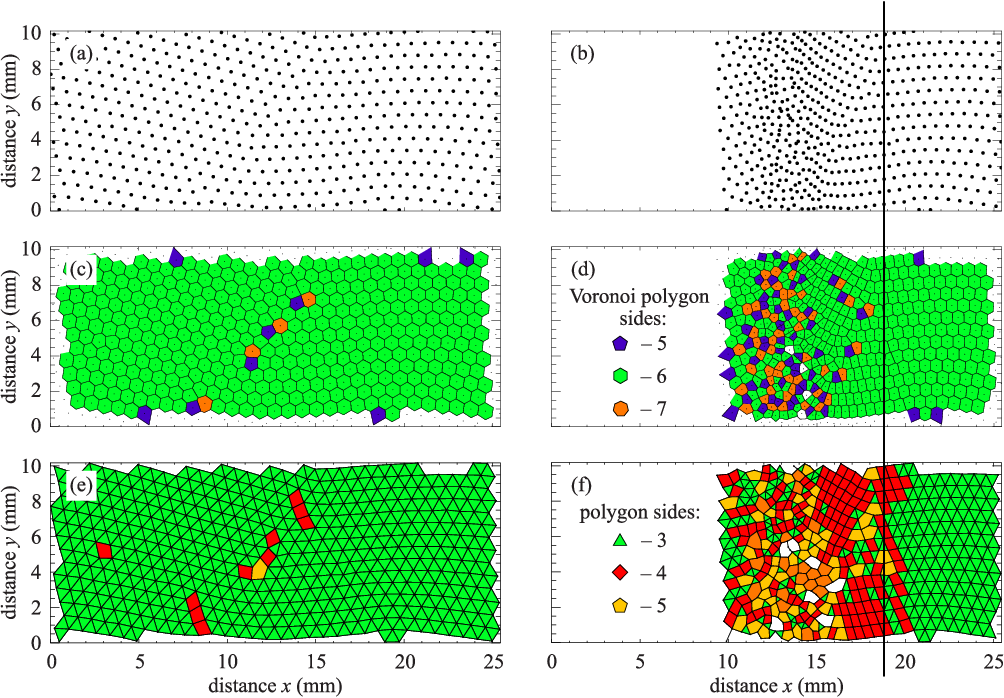}
	\caption{Microscopic structure of the 2D layer of microparticles. Particle positions are shown in (a) and~(b). In the region of interest (ROI) shown here, the exciter wire moved from left to right. The left column is for an early time when the shock had not yet disturbed the ROI, while the right column at 248~ms later had a shock at the location marked with a black line. Voronoi diagrams in (c) and~(d), and polygons in (e) and~(f), reveal regions of order and disorder differently. In Voronoi diagrams, a topological defect corresponds to a non-6-fold coordination, while in a polygon construction map, a geometric defect is recognized as a polygon with more than four sides. The polygon construction map sensitively reveals a non-hexagonal arrangement, near the shock, where there is an abundance of quadrilateral polygons. These data are from a run, corresponding to Fig.~2 in Ref.~\cite{kananovich2020experimental}, with speeds of 101.6~ms/s for the exciter and 117.5~mm/s for the shock. In the postshock for this run, the crystal has clearly melted, as indicated by the considerable disorder that is typical of a liquid. The snapshots shown here are specific frames from videos r\textunderscore A.mp4, v\textunderscore A.mp4, and g\textunderscore A.mp4  in the Supplemental Material~\cite{Note1}.}
	\label{fWidthShockedUnshocked1} 
\end{figure*}

Two methods of structural analysis were used for this paper: Voronoi analysis~\cite{voronoi1908nouvelles} and the Glaser-Clark polygon analysis method~\cite{glaser1990statistical,glaser1992melting}. Both methods have been used in the literature to characterize defects, for the melting of a 2D crystalline lattice. Both of these methods start with particle position measurements as the input for a Delaunay triangulation calculation, which identifies bonds as line segments connecting nearby particles. In an ideal 2D crystal under equilibrium conditions at zero temperature, these bonds will be arranged as equilateral triangles, i.e., with six-fold symmetry. However, in actual conditions, they will not be perfectly equilateral.

The Voronoi method has been used to study 2D melting with experimental data since at least the 1987 colloid experiments of Ref.~\cite{murray1987experimental}. The Voronoi algorithm is to form a polygon having its edges being the perpendicular bisectors of the bonds. In an equilibrium state of a 2D crystal, these Voronoi polygons will be hexagons, which is to say the local geometry has six-fold coordination. Five- and seven-sided polygons, and any other polygon that is not six-fold, are identified as a disclination, which is a defect. This kind of defect analysis has an essentially binary outcome: the vertex at the center of a polygon is either six-fold or it is not. Sufficient displacements of particles from the ground-state arrangement can result in a pair of polygons changing from hexagons to a five-seven pair of Voronoi cells. Sometimes only a tiny displacement of a single particle can result in such a change, from non-defect to defect. This binary classification, defect vs non-defect, is a limitation of the Voronoi method.

The Glaser-Clark polygon analysis method, which was developed in 1990~\cite{glaser1990statistical}, has a distinctive difference compared to the Voronoi approach because it does not have a simple binary outcome of defects vs non-defects. Instead, Glaser-Clark  polygons have various orders: triangular, quadrilateral, pentagonal, or even higher order polygons. A triangular polygon is non-defective. A quadrilateral is the least severe defect, followed by a pentagon and so on for more severe defects~\cite{radzvilavcius2012geometrical,ruhunusiri2011polygon}. This gradation of defects is an advantage of the Glaser-Clark method. The algorithm for producing these polygons from the Delaunay triangulation is to remove bonds if their angles deviate sufficiently from an equilateral shape. The criterion we will use, as in Refs~\cite{ruhunusiri2011polygon,glaser1990statistical,glaser1992melting}, is to remove bonds opposite an unusually large angle of 75\textdegree. We used the source code for Glaser-Clark polygon analysis provided in the Supplementary Material of Ref.~\cite{ruhunusiri2011polygon}. Despite its advantages, this method has not been used as much as Voronoi analysis in the literature. Glaser and Clark originally demonstrated it using simulation data for a 2D lattice under steady conditions~\cite{glaser1990statistical,glaser1992melting}; it was then used with experimental data to characterize the sudden melting of a 2D dusty plasma crystal~\cite{ruhunusiri2011polygon,ruhunusiri2014investigation}, the melting process in 2D Yukawa systems~\cite{radzvilavcius2012geometrical}, and structure and dynamics in screened 2D dipole systems~\cite{aldakul2020melting}.

\section{Results}

\begin{figure}[ht]
	\centering
	\includegraphics[width=1.0\columnwidth]{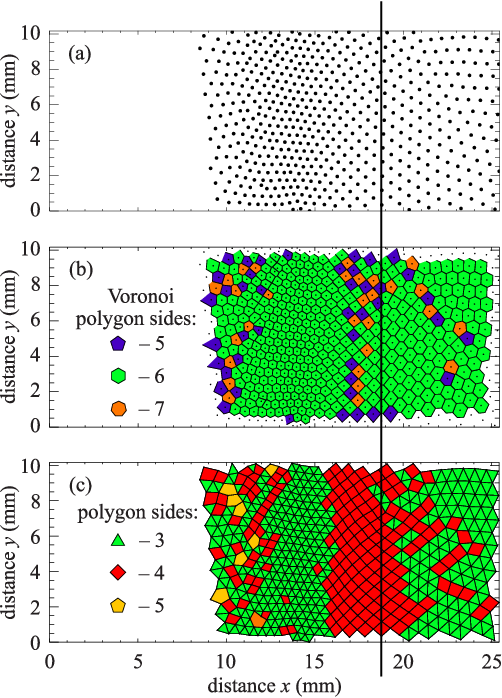}
	\caption{Microscopic structure, with data from a run with a slower speed of 88.9~mm/s for the exciter and 110.5~mm/s for the shock. In this run, there was again an abundance of quadrilaterals near the shock. In the postshock, the rearrangement relaxed more nearly toward a hexagonal lattice, meaning that there is no strong indication of melting. These snapshots are from videos r\textunderscore C.mp4, v\textunderscore C.mp4, and g\textunderscore C.mp4 in the Supplemental Material~\cite{Note1}.}
	\label{fWidthShocked2} 
\end{figure}

\begin{figure}[ht]
	\centering
	\includegraphics[width=1.0\columnwidth]{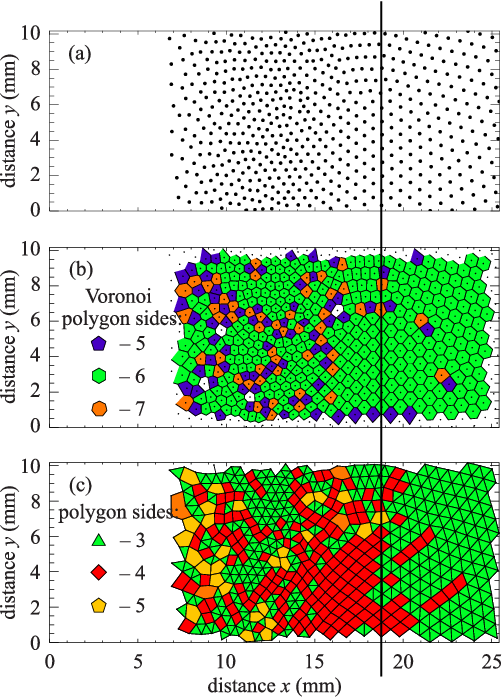}
	\caption{Microscopic structure, with data from a run with an even slower speed of 63.5 mm/s~for the exciter and 83.8~mm/s for the shock. As in the other runs, quadrilaterals were abundant near the shock. The disorder in the postshock was more than in Fig.~\ref{fWidthShocked2}, and almost as much as in Fig.~\ref{fWidthShockedUnshocked1}. We note that the speed of the shock, by itself, does not determine the outcome of whether the postshock is melted. These snapshots are from videos r\textunderscore E.mp4, v\textunderscore E.mp4, and g\textunderscore E.mp4 in the Supplemental Material~\cite{Note1}.}
	\label{fWidthShocked3} 
\end{figure}

In Figs.~\ref{fWidthShockedUnshocked1}-\ref{fWidthShocked3} we present 2D maps of the microstructure. The particle positions shown in the upper rows of these figures were the input to a Delaunay triangulation calculation, for preparing the Voronoi polygons in the middle rows and the Glaser-Clark polygons in the lower rows. These three figures are for different runs, each with a different speed for the exciter wire, which moved in the positive $x$~direction. In Fig.~\ref{fWidthShockedUnshocked1} we also provide an example of the conditions of the microparticle cloud before it was disturbed by the movement of the exciter.  In the Supplemental Material~\cite{Note1} we provide videos of the particle positions, Voronoi cells and Glaser-Clark polygons; these videos are for six experimental runs, including the three runs shown in the snapshots of Figs.~\ref{fWidthShockedUnshocked1}-\ref{fWidthShocked3}.

The shock has a peak density, at a location marked as a solid line in Figs.~\ref{fWidthShockedUnshocked1}-\ref{fWidthShocked3}. We refer to that point as the shock position. Immediately in front of the shock position there is  a localized high-gradient region, which we analyzed in~Ref.~\cite{kananovich2021shock} to quantify the  finite thickness of the shock.

Our chief result is the finding that  microstructures rearrange from hexagonal to quadrilateral within a shock.  Such a quadrilateral structure is seen in Figs.~\ref{fWidthShockedUnshocked1}-\ref{fWidthShocked3}, and in the videos in the Supplemental Material~\cite{Note1}.

The quadrilateral structure is most often observed at locations behind the shock position. In some instances, it also extends slightly ahead of the shock position, as seen in the Glaser-Clark polygon maps. This quadrilateral region is located between what we call the preshock and postshock. The preshock region is located at the most positive $x$-coordinate.  Here, we distinguish the quadrilateral region and the postshock not according to a numerical criterion, but by our qualitative observation of the type of disorder seen in the local microstructure. In our experiment, the quadrilateral region typically had a thickness of about 4~mm. 

An additional result is a considerable increase in defects in the post-shock region.  The defect fraction is quantified as the number of non-six-fold Voronoi cells, divided by the total number of cells within a given region. Values of this defect fraction are printed on each video frame in the Voronoi-map movies in Supplemental Material~\cite{Note1}, and they are summarized in Table I. In all six runs the defect fraction behind the shock was mainly greater than the 25\% level considered to be a liquid in the melting experiment of Ref.~\cite{feng2008solid}. For comparison, the defect fraction was only 3\% to 14\% in the unperturbed condition, before the shock arrived in the ROI.

\begin{table*}[]
	\caption{Runs from the experiment of Ref.~\cite{kananovich2020experimental}. The defect fraction is calculated as the number of non-six-fold Voronoi cells divided by the total number of Voronoi cells, in a spatial region at least 1.25~mm behind the point of highest density. Values for this defect fraction are reported as a typical range over the course of time; values for specific video frames are printed in the videos in the Supplemental Material~\cite{Note1}. The unperturbed conditions were obtained early in each run before the microparticles in the ROI were distrubed noticeably, while the conditions behind the shock are for video frames when the point of maximum density had reached at least the farthest 70\% of the ROI. The Exciter speed and Mach number data are the same as in Fig.~5~of~Ref.~\cite{kananovich2020experimental}. For comparison, we note that \citeauthor{feng2008solid}~\cite{feng2008solid} found in their melting experiment that a liquid typically has  a defect fraction greater than approximately 0.20.}
	\begin{ruledtabular}
		\begin{tabular}{p{0.4in} p{0.7in} p{0.7in} p{1.0in} p{1.0in}  p{1.0in} }
			
		Run	& $\mathrm{Exciter}\,\mathrm{speed}$, mm/s  & $\mathrm{Exciter}\,\mathrm{speed} / a \omega_p$  & Exciter\:Mach\:Number & Defect\:fraction: unperturbed & Defect\:fraction: behind the shock \\ \hline
			                                                                                                                
			A  & 101.6       & 4.9        & 5.36        & 0.05 - 0.08   & 0.41 – 0.47     \\ 
			B  & 101.6       & 4.9        & 5.36        & 0.02 – 0.06   & 0.47 – 0.59      \\
			C  & 88.9        & 4.3        & 4.69        & 0.07 – 0.08   & 0.17 – 0.35      \\
			D  & 76.2        & 3.7        & 4.02        & 0.05 – 0.08   & 0.26 – 0.37      \\ 
			E  & 63.5        & 3.1        & 3.35        & 0.03 – 0.10   & 0.33 – 0.42      \\ 
			F  & 50.8        & 2.5        & 2.68        & 0.10 – 0.14   & 0.41 – 0.48      \\  
			 
		\end{tabular}
	\end{ruledtabular}
	\label{tWidthParam}
\end{table*}

\begin{figure}[ht]
	\centering
	\includegraphics[width=1.0\columnwidth]{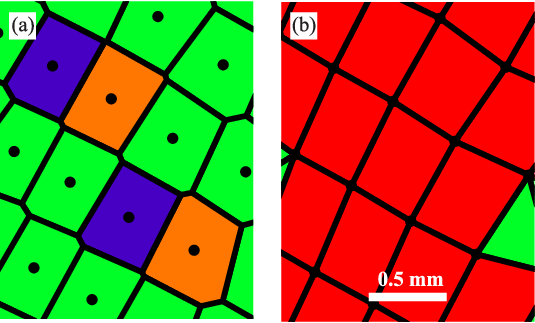}
	\caption{Magnification of Fig.~\ref{fWidthShockedUnshocked1}, showing details in the microscopic structure at a location near $x = 16.4$~mm and $y = 7.7$~mm. Under the compressed condition near the shock, six-fold Voronoi cells (a) have sides that are very unequal in length, with some sides that are so short that it is not very meaningful to say that the coordination is six-fold. Under these nonequilibrium conditions, we find that the polygons in (b) better indicate the nearly rectangular arrangement of particles.}
	\label{fZoom} 
\end{figure}

Glaser-Clark polygon maps are more useful than Voronoi analysis, for the purpose of identifying a quadrilateral structure. A limitation of Voronoi polygons is that they are most likely to have 5, 6 or 7~sides, while very seldomly do they have 4~sides. The Voronoi mapping struggles to map onto a quadrilateral arrangement particles, doing so only by making some of the polygon sides extremely short, as can be seen in the magnified view in~Fig.~\ref{fZoom}.  The  Voronoi cells in~Fig.~\ref{fZoom}~(a) might at a glance appear to be quadrilateral, but a close inspection shows that they are mostly 6-sided, along with some that are 5 or 7-sided, but always with some sides that are extremely short.  

\section{Discussion}

An analysis method that had not previously been applied to 2D shocks allowed us to find that within a shock, a hexagonal microstructure rearranged into a quadrilateral microstructure that had not be reported before. This analysis method was the Glaser-Clark polygon analysis, which we compared to the more common Voronoi analysis method. The data that we analyzed were from runs of the 2D dusty plasma shock experiment of Ref.~\cite{kananovich2020experimental}. An advantage of the Glaser-Clark polygons that we found was that it easily reveals quadrilateral structures, while Voronoi polygons are not well suited for this purpose, as they are mostly 5, 6 or 7-sided. Another advantage of Glaser-Clark polygon anlaysis is that a gradually increasing disorder in the microstructure is quantified by a gradual increase in the number of sides of the polygons~\cite{ruhunusiri2011polygon}.

The quadrilateral structure that we observed is generally not otherwise seen, in  a single layer of identical particles interacting isotropically. A triangular lattice with six-fold symmetry is the only stable lattice in this case. As an example, a square lattice configuration is not stable in equilibrium, as the total interparticle potential energy can be diminished by alternate rows slipping so that the lattice becomes triangular.

We believe that it may be significant that the quadrilateral structure was observed at locations where the conditions were strongly nonequilibrium. Shocks are certainly nonequilibrium, especially near the location of greatest compression, and it was near this location where we observed quadrilateral structures. 

There are previous dusty plasma experiments, done under nearly 2D conditions, where quadrilateral microstructures have been observed under steady conditions. These experiments, which did not involve shocks, fall in two categories: those with an externally imposed pattern, and those with out-of-plane particles. An externally imposed pattern was applied to a magnetized plasma, when a metal mesh with an 0.56~mm square pattern was inserted as a boundary in the Magnetized Dusty Plasma Experiment (MDPX) device~\cite{thomas2015quasi,thomas2015observations,hall2018methods}, causing the same square pattern to appear in the arrangement of dust particles.  This effect can be attributed to the external influence of the mesh, which persisted throughout the volume due to the magnetization of electrons and ions. For unmagnetized plasmas, we note two experiments that involved out-of-plane positions of some particles. In both of these experiments authors attributed their observations of quadrilateral microstructures to non-isotropic interparticle interactions: ion wakes beneath the upper particles that disturbed the lower particles.  In 2019 \citeauthor{huang2019wave}~\cite{huang2019wave} reported an experiment with  a binary mixture, and they observed distinctive regions with four-fold symmetry rather than the six-fold symmetry that is common in a~2D~crystal with monodisperse microparticles. In 2022 \citeauthor{singh2022square}~\cite{singh2022square} used monodisperse particles and attained a quadrilateral microstructure by causing the microparticle layer to buckle, meaning that the horizontal repulsion between microparticles overcame the confining forces in the vertical direction.

Shocks in a 2D layer have been studied previously in dusty plasma experiments \cite{samsonov2004shock,oxtoby2013ideal} and in molecular dynamics simulations~\cite{marciante2017thermodynamic,lin2019pressure,lin2020universal,sun2021evolution,ding2021head,sun2021shock,qiu2021observation}. The latter mimicked dusty plasmas by tracking the motion of identical particles interacting with a repulsive Yukawa potential, but without capturing all the effects in the experiment such as out-of-plane displacements and finite dispersion of particle size. In none of the experiment or simulation papers that we reviewed did the authors comment upon a quadrilateral microstructure. Nevertheless,  we see indications of quadrilateral shapes in both the experiment in Fig.~1 of Ref.~\cite{oxtoby2013ideal} and in Fig.~1(a) of the Supplemental Material of Ref.~\cite{marciante2017thermodynamic}. These quadrilateral structures appear in the vicinity of the peak density of a blast-wave shock in the experiment of Ref.~\cite{oxtoby2013ideal} but on the pre-shock side in the simulation of a steadily-driven shock in  Ref.~\cite{marciante2017thermodynamic}.

Besides our primary result of finding a quadrilateral microstructure, we also discussed the production of defects and melting. The defect fraction was considerably higher behind the shock as compared to the unperturbed condition in six runs that were analyzed. In fact, behind the shock the defect fraction was generally so high as to be consistent with melting. We did not observe a distinctive trend for the defect fraction to increase with exciter speed. We note that all of the exciter speeds in the experiment were at Mach numbers considerably higher than in recent simulations~\cite{qiu2021observation} and \cite{sun2021shock}  where the onset of shock-induced melting was studied. Some factors that could affect the defect fraction behind the shock include the presence, in the earlier unperturbed structure, of domain walls and other defects, which vary from one experimental run to another. One more factor that can vary from one run to another is the direction of the lattice alignment in the unperturbed condition. These conditions for shocks in a solid differ from those of gasdynamic shocks, where the unshocked region has no microscopic structure or favored directions. 

\begin{acknowledgments}
This work was supported by the Army Research Office under MURI Grant No. W911NF-18-1-0240, the United States Department of Energy under Grant No. DE-SC0014566 and NASA/JPL RSA Nos. 1689926, and the National Science Foundation under Grant No. PHY-1740379.
\end{acknowledgments}


\begin{thebibliography}{45}
	\expandafter\ifx\csname natexlab\endcsname\relax\def\natexlab#1{#1}\fi
	\expandafter\ifx\csname bibnamefont\endcsname\relax
	\def\bibnamefont#1{#1}\fi
	\expandafter\ifx\csname bibfnamefont\endcsname\relax
	\def\bibfnamefont#1{#1}\fi
	\expandafter\ifx\csname citenamefont\endcsname\relax
	\def\citenamefont#1{#1}\fi
	\expandafter\ifx\csname url\endcsname\relax
	\def\url#1{\texttt{#1}}\fi
	\expandafter\ifx\csname urlprefix\endcsname\relax\def\urlprefix{URL }\fi
	\providecommand{\bibinfo}[2]{#2}
	\providecommand{\eprint}[2][]{\url{#2}}
	
	\bibitem[{\citenamefont{Goree}(2008)}]{goree2008fundamentals}
	\bibinfo{author}{\bibfnamefont{J.}~\bibnamefont{Goree}},
	\emph{\bibinfo{title}{Fundamentals of Dusty Plasmas}}
	(\bibinfo{publisher}{Wiley-VCH}, \bibinfo{year}{2008}),
	chap.~\bibinfo{chapter}{6}, pp. \bibinfo{pages}{157--206}.
	
	\bibitem[{\citenamefont{Melzer}(2019)}]{melzer2019physics}
	\bibinfo{author}{\bibfnamefont{A.}~\bibnamefont{Melzer}},
	\emph{\bibinfo{title}{Physics of Dusty Plasmas}}, vol. \bibinfo{volume}{962}
	of \emph{\bibinfo{series}{Lecture Notes in Physics}}
	(\bibinfo{publisher}{Springer International Publishing},
	\bibinfo{address}{Cham}, \bibinfo{year}{2019}).
	
	\bibitem[{\citenamefont{Samsonov et~al.}(1999)\citenamefont{Samsonov, Goree,
			Ma, Bhattacharjee, Thomas, and Morfill}}]{samsonov1999mach}
	\bibinfo{author}{\bibfnamefont{D.}~\bibnamefont{Samsonov}},
	\bibinfo{author}{\bibfnamefont{J.}~\bibnamefont{Goree}},
	\bibinfo{author}{\bibfnamefont{Z.~W.} \bibnamefont{Ma}},
	\bibinfo{author}{\bibfnamefont{A.}~\bibnamefont{Bhattacharjee}},
	\bibinfo{author}{\bibfnamefont{H.~M.} \bibnamefont{Thomas}},
	\bibnamefont{and} \bibinfo{author}{\bibfnamefont{G.~E.}
		\bibnamefont{Morfill}}, \bibinfo{journal}{Phys. Rev. Lett.}
	\textbf{\bibinfo{volume}{83}}, \bibinfo{pages}{3649} (\bibinfo{year}{1999}).
	
	\bibitem[{\citenamefont{Samsonov et~al.}(2000)\citenamefont{Samsonov, Goree,
			Thomas, and Morfill}}]{samsonov2000mach}
	\bibinfo{author}{\bibfnamefont{D.}~\bibnamefont{Samsonov}},
	\bibinfo{author}{\bibfnamefont{J.}~\bibnamefont{Goree}},
	\bibinfo{author}{\bibfnamefont{H.~M.} \bibnamefont{Thomas}},
	\bibnamefont{and} \bibinfo{author}{\bibfnamefont{G.~E.}
		\bibnamefont{Morfill}}, \bibinfo{journal}{Phys. Rev. E}
	\textbf{\bibinfo{volume}{61}}, \bibinfo{pages}{5557} (\bibinfo{year}{2000}).
	
	\bibitem[{\citenamefont{Samsonov et~al.}(2004)\citenamefont{Samsonov, Zhdanov,
			Quinn, Popel, and Morfill}}]{samsonov2004shock}
	\bibinfo{author}{\bibfnamefont{D.}~\bibnamefont{Samsonov}},
	\bibinfo{author}{\bibfnamefont{S.~K.} \bibnamefont{Zhdanov}},
	\bibinfo{author}{\bibfnamefont{R.~A.} \bibnamefont{Quinn}},
	\bibinfo{author}{\bibfnamefont{S.~I.} \bibnamefont{Popel}}, \bibnamefont{and}
	\bibinfo{author}{\bibfnamefont{G.~E.} \bibnamefont{Morfill}},
	\bibinfo{journal}{Phys. Rev. Lett.} \textbf{\bibinfo{volume}{92}},
	\bibinfo{pages}{255004} (\bibinfo{year}{2004}).
	
	\bibitem[{\citenamefont{Cou\"edel et~al.}(2012)\citenamefont{Cou\"edel,
			Samsonov, Durniak, Zhdanov, Thomas, Morfill, and Arnas}}]{couedel2012three}
	\bibinfo{author}{\bibfnamefont{L.}~\bibnamefont{Cou\"edel}},
	\bibinfo{author}{\bibfnamefont{D.}~\bibnamefont{Samsonov}},
	\bibinfo{author}{\bibfnamefont{C.}~\bibnamefont{Durniak}},
	\bibinfo{author}{\bibfnamefont{S.}~\bibnamefont{Zhdanov}},
	\bibinfo{author}{\bibfnamefont{H.~M.} \bibnamefont{Thomas}},
	\bibinfo{author}{\bibfnamefont{G.~E.} \bibnamefont{Morfill}},
	\bibnamefont{and} \bibinfo{author}{\bibfnamefont{C.}~\bibnamefont{Arnas}},
	\bibinfo{journal}{Phys. Rev. Lett.} \textbf{\bibinfo{volume}{109}},
	\bibinfo{pages}{175001} (\bibinfo{year}{2012}).
	
	\bibitem[{\citenamefont{Lin et~al.}(2019)\citenamefont{Lin, Murillo, and
			Feng}}]{lin2019pressure}
	\bibinfo{author}{\bibfnamefont{W.}~\bibnamefont{Lin}},
	\bibinfo{author}{\bibfnamefont{M.~S.} \bibnamefont{Murillo}},
	\bibnamefont{and} \bibinfo{author}{\bibfnamefont{Y.}~\bibnamefont{Feng}},
	\bibinfo{journal}{Phys. Rev. E} \textbf{\bibinfo{volume}{100}},
	\bibinfo{pages}{043203} (\bibinfo{year}{2019}).
	
	\bibitem[{\citenamefont{Kananovich and
			Goree}(2020{\natexlab{a}})}]{kananovich2020shocksa}
	\bibinfo{author}{\bibfnamefont{A.}~\bibnamefont{Kananovich}} \bibnamefont{and}
	\bibinfo{author}{\bibfnamefont{J.}~\bibnamefont{Goree}},
	\bibinfo{journal}{Phys. Plasmas} \textbf{\bibinfo{volume}{27}},
	\bibinfo{pages}{113704} (\bibinfo{year}{2020}{\natexlab{a}}).
	
	\bibitem[{\citenamefont{Kananovich and
			Goree}(2020{\natexlab{b}})}]{kananovich2020experimental}
	\bibinfo{author}{\bibfnamefont{A.}~\bibnamefont{Kananovich}} \bibnamefont{and}
	\bibinfo{author}{\bibfnamefont{J.}~\bibnamefont{Goree}},
	\bibinfo{journal}{Phys. Rev. E} \textbf{\bibinfo{volume}{101}},
	\bibinfo{pages}{043211} (\bibinfo{year}{2020}{\natexlab{b}}).
	
	\bibitem[{\citenamefont{Lin et~al.}(2020)\citenamefont{Lin, Murillo, and
			Feng}}]{lin2020universal}
	\bibinfo{author}{\bibfnamefont{W.}~\bibnamefont{Lin}},
	\bibinfo{author}{\bibfnamefont{M.~S.} \bibnamefont{Murillo}},
	\bibnamefont{and} \bibinfo{author}{\bibfnamefont{Y.}~\bibnamefont{Feng}},
	\bibinfo{journal}{Phys. Rev. E} \textbf{\bibinfo{volume}{101}},
	\bibinfo{pages}{013203} (\bibinfo{year}{2020}).
	
	\bibitem[{\citenamefont{Tao and Duan}(2020)}]{tao2020effects}
	\bibinfo{author}{\bibfnamefont{L.-L.} \bibnamefont{Tao}} \bibnamefont{and}
	\bibinfo{author}{\bibfnamefont{W.-S.} \bibnamefont{Duan}},
	\bibinfo{journal}{Chin. J. Phys.} \textbf{\bibinfo{volume}{68}},
	\bibinfo{pages}{950} (\bibinfo{year}{2020}).
	
	\bibitem[{\citenamefont{Sun and Feng}(2021)}]{sun2021shock}
	\bibinfo{author}{\bibfnamefont{T.}~\bibnamefont{Sun}} \bibnamefont{and}
	\bibinfo{author}{\bibfnamefont{Y.}~\bibnamefont{Feng}},
	\bibinfo{journal}{Phys. Plasmas} \textbf{\bibinfo{volume}{28}},
	\bibinfo{pages}{063702} (\bibinfo{year}{2021}).
	
	\bibitem[{\citenamefont{Qiu et~al.}(2021)\citenamefont{Qiu, Sun, and
			Feng}}]{qiu2021observation}
	\bibinfo{author}{\bibfnamefont{P.}~\bibnamefont{Qiu}},
	\bibinfo{author}{\bibfnamefont{T.}~\bibnamefont{Sun}}, \bibnamefont{and}
	\bibinfo{author}{\bibfnamefont{Y.}~\bibnamefont{Feng}},
	\bibinfo{journal}{Phys. Plasmas} \textbf{\bibinfo{volume}{28}},
	\bibinfo{pages}{113702} (\bibinfo{year}{2021}).
	
	\bibitem[{\citenamefont{Kananovich and Goree}(2021)}]{kananovich2021shock}
	\bibinfo{author}{\bibfnamefont{A.}~\bibnamefont{Kananovich}} \bibnamefont{and}
	\bibinfo{author}{\bibfnamefont{J.}~\bibnamefont{Goree}},
	\bibinfo{journal}{Phys. Rev. E} \textbf{\bibinfo{volume}{104}},
	\bibinfo{pages}{055201} (\bibinfo{year}{2021}).
	
	\bibitem[{\citenamefont{Sun et~al.}(2021)\citenamefont{Sun, Murillo, and
			Feng}}]{sun2021evolution}
	\bibinfo{author}{\bibfnamefont{T.}~\bibnamefont{Sun}},
	\bibinfo{author}{\bibfnamefont{M.~S.} \bibnamefont{Murillo}},
	\bibnamefont{and} \bibinfo{author}{\bibfnamefont{Y.}~\bibnamefont{Feng}},
	\bibinfo{journal}{Phys. Plasmas} \textbf{\bibinfo{volume}{28}},
	\bibinfo{pages}{103703} (\bibinfo{year}{2021}).
	
	\bibitem[{\citenamefont{Ding et~al.}(2021)\citenamefont{Ding, Lu, Sun, Murillo,
			and Feng}}]{ding2021head}
	\bibinfo{author}{\bibfnamefont{X.}~\bibnamefont{Ding}},
	\bibinfo{author}{\bibfnamefont{S.}~\bibnamefont{Lu}},
	\bibinfo{author}{\bibfnamefont{T.}~\bibnamefont{Sun}},
	\bibinfo{author}{\bibfnamefont{M.~S.} \bibnamefont{Murillo}},
	\bibnamefont{and} \bibinfo{author}{\bibfnamefont{Y.}~\bibnamefont{Feng}},
	\bibinfo{journal}{Phys. Rev. E} \textbf{\bibinfo{volume}{103}},
	\bibinfo{pages}{013202} (\bibinfo{year}{2021}).
	
	\bibitem[{\citenamefont{Gou et~al.}(2021)\citenamefont{Gou, An, and
			Duan}}]{gou2021effects}
	\bibinfo{author}{\bibfnamefont{X.-Q.} \bibnamefont{Gou}},
	\bibinfo{author}{\bibfnamefont{K.-H.} \bibnamefont{An}}, \bibnamefont{and}
	\bibinfo{author}{\bibfnamefont{W.-S.} \bibnamefont{Duan}},
	\bibinfo{journal}{Braz. J. Phys.} \textbf{\bibinfo{volume}{51}},
	\bibinfo{pages}{1346} (\bibinfo{year}{2021}).
	
	\bibitem[{\citenamefont{Li and Duan}(2021)}]{li2021weak}
	\bibinfo{author}{\bibfnamefont{Z.-Z.} \bibnamefont{Li}} \bibnamefont{and}
	\bibinfo{author}{\bibfnamefont{W.-S.} \bibnamefont{Duan}},
	\bibinfo{journal}{Phys. Plasmas} \textbf{\bibinfo{volume}{28}}
	(\bibinfo{year}{2021}).
	
	\bibitem[{\citenamefont{Tao et~al.}(2021)\citenamefont{Tao, Wei, Liu, Zhang,
			and Duan}}]{tao2021effect}
	\bibinfo{author}{\bibfnamefont{L.-L.} \bibnamefont{Tao}},
	\bibinfo{author}{\bibfnamefont{L.}~\bibnamefont{Wei}},
	\bibinfo{author}{\bibfnamefont{B.}~\bibnamefont{Liu}},
	\bibinfo{author}{\bibfnamefont{H.}~\bibnamefont{Zhang}}, \bibnamefont{and}
	\bibinfo{author}{\bibfnamefont{W.-S.} \bibnamefont{Duan}},
	\bibinfo{journal}{Pramana-J. Phys.} \textbf{\bibinfo{volume}{95}}
	(\bibinfo{year}{2021}).
	
	\bibitem[{\citenamefont{Zhang et~al.}(2022)\citenamefont{Zhang, Xia, and
			Liu}}]{zhang2022structural}
	\bibinfo{author}{\bibfnamefont{X.}~\bibnamefont{Zhang}},
	\bibinfo{author}{\bibfnamefont{J.}~\bibnamefont{Xia}}, \bibnamefont{and}
	\bibinfo{author}{\bibfnamefont{X.}~\bibnamefont{Liu}},
	\bibinfo{journal}{Phys. Rev. B} \textbf{\bibinfo{volume}{105}}
	(\bibinfo{year}{2022}).
	
	\bibitem[{\citenamefont{Samsonov and Morfill}(2008)}]{samsonov2008high}
	\bibinfo{author}{\bibfnamefont{D.}~\bibnamefont{Samsonov}} \bibnamefont{and}
	\bibinfo{author}{\bibfnamefont{G.}~\bibnamefont{Morfill}},
	\bibinfo{journal}{IEEE Trans. Plasma Sci.} \textbf{\bibinfo{volume}{36}},
	\bibinfo{pages}{1020} (\bibinfo{year}{2008}).
	
	\bibitem[{\citenamefont{Chaikin and Lubensky}(1995)}]{chaikin1995principles}
	\bibinfo{author}{\bibfnamefont{P.~M.} \bibnamefont{Chaikin}} \bibnamefont{and}
	\bibinfo{author}{\bibfnamefont{T.~C.} \bibnamefont{Lubensky}},
	\emph{\bibinfo{title}{Principles of Condensed Matter Physics}}
	(\bibinfo{publisher}{Cambridge University Press}, \bibinfo{address}{New
		York}, \bibinfo{year}{1995}), pp. \bibinfo{pages}{50--53}.
	
	\bibitem[{\citenamefont{Huang et~al.}(2019)\citenamefont{Huang, Ivlev, Nosenko,
			Lin, and Du}}]{huang2019wave}
	\bibinfo{author}{\bibfnamefont{H.}~\bibnamefont{Huang}},
	\bibinfo{author}{\bibfnamefont{A.~V.} \bibnamefont{Ivlev}},
	\bibinfo{author}{\bibfnamefont{V.}~\bibnamefont{Nosenko}},
	\bibinfo{author}{\bibfnamefont{Y.-F.} \bibnamefont{Lin}}, \bibnamefont{and}
	\bibinfo{author}{\bibfnamefont{C.-R.} \bibnamefont{Du}},
	\bibinfo{journal}{Phys. Plasmas} \textbf{\bibinfo{volume}{26}},
	\bibinfo{pages}{013702} (\bibinfo{year}{2019}).
	
	\bibitem[{\citenamefont{Zampetaki et~al.}(2020)\citenamefont{Zampetaki, Huang,
			Du, L\"owen, and Ivlev}}]{zampetaki2020buckling}
	\bibinfo{author}{\bibfnamefont{A.~V.} \bibnamefont{Zampetaki}},
	\bibinfo{author}{\bibfnamefont{H.}~\bibnamefont{Huang}},
	\bibinfo{author}{\bibfnamefont{C.-R.} \bibnamefont{Du}},
	\bibinfo{author}{\bibfnamefont{H.}~\bibnamefont{L\"owen}}, \bibnamefont{and}
	\bibinfo{author}{\bibfnamefont{A.~V.} \bibnamefont{Ivlev}},
	\bibinfo{journal}{Phys. Rev. E} \textbf{\bibinfo{volume}{102}},
	\bibinfo{pages}{043204} (\bibinfo{year}{2020}).
	
	\bibitem[{\citenamefont{Singh et~al.}(2022)\citenamefont{Singh, Bandyopadhyay,
			Kumar, and Sen}}]{singh2022square}
	\bibinfo{author}{\bibfnamefont{S.}~\bibnamefont{Singh}},
	\bibinfo{author}{\bibfnamefont{P.}~\bibnamefont{Bandyopadhyay}},
	\bibinfo{author}{\bibfnamefont{K.}~\bibnamefont{Kumar}}, \bibnamefont{and}
	\bibinfo{author}{\bibfnamefont{A.}~\bibnamefont{Sen}},
	\bibinfo{journal}{Phys. Rev. Lett.} \textbf{\bibinfo{volume}{129}},
	\bibinfo{pages}{115003} (\bibinfo{year}{2022}).
	
	\bibitem[{\citenamefont{Voronoi}(1908)}]{voronoi1908nouvelles}
	\bibinfo{author}{\bibfnamefont{G.}~\bibnamefont{Voronoi}}, \bibinfo{journal}{J.
		Reine Angew. Math.} \textbf{\bibinfo{volume}{1908}}, \bibinfo{pages}{198 }
	(\bibinfo{year}{1908}).
	
	\bibitem[{\citenamefont{Ruhunusiri et~al.}(2011)\citenamefont{Ruhunusiri,
			Goree, Feng, and Liu}}]{ruhunusiri2011polygon}
	\bibinfo{author}{\bibfnamefont{W.~D.~S.} \bibnamefont{Ruhunusiri}},
	\bibinfo{author}{\bibfnamefont{J.}~\bibnamefont{Goree}},
	\bibinfo{author}{\bibfnamefont{Y.}~\bibnamefont{Feng}}, \bibnamefont{and}
	\bibinfo{author}{\bibfnamefont{B.}~\bibnamefont{Liu}},
	\bibinfo{journal}{Phys. Rev. E} \textbf{\bibinfo{volume}{83}},
	\bibinfo{pages}{66402} (\bibinfo{year}{2011}).
	
	\bibitem[{\citenamefont{Glaser and Clark}(1990)}]{glaser1990statistical}
	\bibinfo{author}{\bibfnamefont{M.~A.} \bibnamefont{Glaser}} \bibnamefont{and}
	\bibinfo{author}{\bibfnamefont{N.~A.} \bibnamefont{Clark}},
	\bibinfo{journal}{Phys. Rev. A} \textbf{\bibinfo{volume}{41}},
	\bibinfo{pages}{4585} (\bibinfo{year}{1990}).
	
	\bibitem[{\citenamefont{Glaser and Clark}(1992)}]{glaser1992melting}
	\bibinfo{author}{\bibfnamefont{M.~A.} \bibnamefont{Glaser}} \bibnamefont{and}
	\bibinfo{author}{\bibfnamefont{N.~A.} \bibnamefont{Clark}},
	\emph{\bibinfo{title}{Melting and Liquid Structure in two Dimensions}}
	(\bibinfo{publisher}{Wiley}, \bibinfo{year}{1992}), pp.
	\bibinfo{pages}{543--709}.
	
	\bibitem[{\citenamefont{Ruhunusiri}(2014)}]{ruhunusiri2014investigation}
	\bibinfo{author}{\bibfnamefont{W.~D.~S.} \bibnamefont{Ruhunusiri}}, Ph.D.
	thesis, \bibinfo{school}{the University of Iowa} (\bibinfo{year}{2014}).
	
	\bibitem[{\citenamefont{Chaubey et~al.}(2021)\citenamefont{Chaubey, Goree,
			Lanham, and Kushner}}]{chaubey2021positive}
	\bibinfo{author}{\bibfnamefont{N.}~\bibnamefont{Chaubey}},
	\bibinfo{author}{\bibfnamefont{J.}~\bibnamefont{Goree}},
	\bibinfo{author}{\bibfnamefont{S.~J.} \bibnamefont{Lanham}},
	\bibnamefont{and} \bibinfo{author}{\bibfnamefont{M.~J.}
		\bibnamefont{Kushner}}, \bibinfo{journal}{Phys. Plasmas}
	\textbf{\bibinfo{volume}{28}}, \bibinfo{pages}{103702}
	(\bibinfo{year}{2021}).
	
	\bibitem[{Note1()}]{Note1}
	Note1, \bibinfo{note}{see Supplemental Material at [URL will be inserted by
		publisher]}.
	
	\bibitem[{\citenamefont{Feng et~al.}(2007)\citenamefont{Feng, Goree, and
			Liu}}]{feng2007accurate}
	\bibinfo{author}{\bibfnamefont{Y.}~\bibnamefont{Feng}},
	\bibinfo{author}{\bibfnamefont{J.}~\bibnamefont{Goree}}, \bibnamefont{and}
	\bibinfo{author}{\bibfnamefont{B.}~\bibnamefont{Liu}}, \bibinfo{journal}{Rev.
		Sci. Instrum.} \textbf{\bibinfo{volume}{78}}, \bibinfo{eid}{053704}
	(\bibinfo{year}{2007}).
	
	\bibitem[{\citenamefont{Feng et~al.}(2011)\citenamefont{Feng, Goree, and
			Liu}}]{feng2011errors}
	\bibinfo{author}{\bibfnamefont{Y.}~\bibnamefont{Feng}},
	\bibinfo{author}{\bibfnamefont{J.}~\bibnamefont{Goree}}, \bibnamefont{and}
	\bibinfo{author}{\bibfnamefont{B.}~\bibnamefont{Liu}}, \bibinfo{journal}{Rev.
		Sci. Instrum.} \textbf{\bibinfo{volume}{82}}, \bibinfo{eid}{053707}
	(\bibinfo{year}{2011}).
	
	\bibitem[{\citenamefont{Feng et~al.}(2016)\citenamefont{Feng, Goree, Haralson,
			Wong, Kananovich, and Li}}]{feng2016particle}
	\bibinfo{author}{\bibfnamefont{Y.}~\bibnamefont{Feng}},
	\bibinfo{author}{\bibfnamefont{J.}~\bibnamefont{Goree}},
	\bibinfo{author}{\bibfnamefont{Z.}~\bibnamefont{Haralson}},
	\bibinfo{author}{\bibfnamefont{C.-S.} \bibnamefont{Wong}},
	\bibinfo{author}{\bibfnamefont{A.}~\bibnamefont{Kananovich}},
	\bibnamefont{and} \bibinfo{author}{\bibfnamefont{W.}~\bibnamefont{Li}},
	\bibinfo{journal}{J. Plasma Phys.} \textbf{\bibinfo{volume}{82}},
	\bibinfo{pages}{615820303} (\bibinfo{year}{2016}).
	
	\bibitem[{\citenamefont{Schneider et~al.}(2012)\citenamefont{Schneider,
			Rasband, and Eliceiri}}]{schneider2012nih}
	\bibinfo{author}{\bibfnamefont{C.~A.} \bibnamefont{Schneider}},
	\bibinfo{author}{\bibfnamefont{W.~S.} \bibnamefont{Rasband}},
	\bibnamefont{and} \bibinfo{author}{\bibfnamefont{K.~W.}
		\bibnamefont{Eliceiri}}, \bibinfo{journal}{Nat. Methods}
	\textbf{\bibinfo{volume}{9}}, \bibinfo{pages}{671} (\bibinfo{year}{2012}).
	
	\bibitem[{\citenamefont{Murray and Van~Winkle}(1987)}]{murray1987experimental}
	\bibinfo{author}{\bibfnamefont{C.~A.} \bibnamefont{Murray}} \bibnamefont{and}
	\bibinfo{author}{\bibfnamefont{D.~H.} \bibnamefont{Van~Winkle}},
	\bibinfo{journal}{Phys. Rev. Lett.} \textbf{\bibinfo{volume}{58}},
	\bibinfo{pages}{1200} (\bibinfo{year}{1987}).
	
	\bibitem[{\citenamefont{Radzvilav\v{c}ius}(2012)}]{radzvilavcius2012geometrical}
	\bibinfo{author}{\bibfnamefont{A.}~\bibnamefont{Radzvilav\v{c}ius}},
	\bibinfo{journal}{Phys. Rev. E} \textbf{\bibinfo{volume}{86}},
	\bibinfo{pages}{051111} (\bibinfo{year}{2012}).
	
	\bibitem[{\citenamefont{Aldakul et~al.}(2020)\citenamefont{Aldakul, Moldabekov,
			and Ramazanov}}]{aldakul2020melting}
	\bibinfo{author}{\bibfnamefont{Y.~K.} \bibnamefont{Aldakul}},
	\bibinfo{author}{\bibfnamefont{Z.~A.} \bibnamefont{Moldabekov}},
	\bibnamefont{and} \bibinfo{author}{\bibfnamefont{T.~S.}
		\bibnamefont{Ramazanov}}, \bibinfo{journal}{Phys. Rev. E}
	\textbf{\bibinfo{volume}{102}}, \bibinfo{pages}{033205}
	(\bibinfo{year}{2020}).
	
	\bibitem[{\citenamefont{Feng et~al.}(2008)\citenamefont{Feng, Goree, and
			Liu}}]{feng2008solid}
	\bibinfo{author}{\bibfnamefont{Y.}~\bibnamefont{Feng}},
	\bibinfo{author}{\bibfnamefont{J.}~\bibnamefont{Goree}}, \bibnamefont{and}
	\bibinfo{author}{\bibfnamefont{B.}~\bibnamefont{Liu}},
	\bibinfo{journal}{Phys. Rev. Lett.} \textbf{\bibinfo{volume}{100}},
	\bibinfo{pages}{205007} (\bibinfo{year}{2008}).
	
	\bibitem[{\citenamefont{Thomas et~al.}(2015{\natexlab{a}})\citenamefont{Thomas,
			Konopka, Lynch, Adams, LeBlanc, Merlino, and Rosenberg}}]{thomas2015quasi}
	\bibinfo{author}{\bibfnamefont{J.}~\bibnamefont{Thomas},
		\bibfnamefont{Edward}},
	\bibinfo{author}{\bibfnamefont{U.}~\bibnamefont{Konopka}},
	\bibinfo{author}{\bibfnamefont{B.}~\bibnamefont{Lynch}},
	\bibinfo{author}{\bibfnamefont{S.}~\bibnamefont{Adams}},
	\bibinfo{author}{\bibfnamefont{S.}~\bibnamefont{LeBlanc}},
	\bibinfo{author}{\bibfnamefont{R.~L.} \bibnamefont{Merlino}},
	\bibnamefont{and}
	\bibinfo{author}{\bibfnamefont{M.}~\bibnamefont{Rosenberg}},
	\bibinfo{journal}{Phys. Plasmas} \textbf{\bibinfo{volume}{22}}
	(\bibinfo{year}{2015}{\natexlab{a}}), \bibinfo{note}{113708}.
	
	\bibitem[{\citenamefont{Thomas et~al.}(2015{\natexlab{b}})\citenamefont{Thomas,
			Lynch, Konopka, Merlino, and Rosenberg}}]{thomas2015observations}
	\bibinfo{author}{\bibfnamefont{J.}~\bibnamefont{Thomas},
		\bibfnamefont{Edward}},
	\bibinfo{author}{\bibfnamefont{B.}~\bibnamefont{Lynch}},
	\bibinfo{author}{\bibfnamefont{U.}~\bibnamefont{Konopka}},
	\bibinfo{author}{\bibfnamefont{R.~L.} \bibnamefont{Merlino}},
	\bibnamefont{and}
	\bibinfo{author}{\bibfnamefont{M.}~\bibnamefont{Rosenberg}},
	\bibinfo{journal}{Phys. Plasmas} \textbf{\bibinfo{volume}{22}}
	(\bibinfo{year}{2015}{\natexlab{b}}), \bibinfo{note}{030701}.
	
	\bibitem[{\citenamefont{Hall et~al.}(2018)\citenamefont{Hall, Thomas, Avinash,
			Merlino, and Rosenberg}}]{hall2018methods}
	\bibinfo{author}{\bibfnamefont{T.}~\bibnamefont{Hall}},
	\bibinfo{author}{\bibfnamefont{J.}~\bibnamefont{Thomas},
		\bibfnamefont{Edward}},
	\bibinfo{author}{\bibfnamefont{K.}~\bibnamefont{Avinash}},
	\bibinfo{author}{\bibfnamefont{R.}~\bibnamefont{Merlino}}, \bibnamefont{and}
	\bibinfo{author}{\bibfnamefont{M.}~\bibnamefont{Rosenberg}},
	\bibinfo{journal}{Phys. Plasmas} \textbf{\bibinfo{volume}{25}}
	(\bibinfo{year}{2018}), \bibinfo{note}{103702}.
	
	\bibitem[{\citenamefont{Oxtoby et~al.}(2013)\citenamefont{Oxtoby, Griffith,
			Durniak, Ralph, and Samsonov}}]{oxtoby2013ideal}
	\bibinfo{author}{\bibfnamefont{N.~P.} \bibnamefont{Oxtoby}},
	\bibinfo{author}{\bibfnamefont{E.~J.} \bibnamefont{Griffith}},
	\bibinfo{author}{\bibfnamefont{C.}~\bibnamefont{Durniak}},
	\bibinfo{author}{\bibfnamefont{J.~F.} \bibnamefont{Ralph}}, \bibnamefont{and}
	\bibinfo{author}{\bibfnamefont{D.}~\bibnamefont{Samsonov}},
	\bibinfo{journal}{Phys. Rev. Lett.} \textbf{\bibinfo{volume}{111}},
	\bibinfo{pages}{015002} (\bibinfo{year}{2013}).
	
	\bibitem[{\citenamefont{Marciante and
			Murillo}(2017)}]{marciante2017thermodynamic}
	\bibinfo{author}{\bibfnamefont{M.}~\bibnamefont{Marciante}} \bibnamefont{and}
	\bibinfo{author}{\bibfnamefont{M.~S.} \bibnamefont{Murillo}},
	\bibinfo{journal}{Phys. Rev. Lett.} \textbf{\bibinfo{volume}{118}},
	\bibinfo{pages}{025001} (\bibinfo{year}{2017}).
	
\end{thebibliography}
\end{document}